\documentclass[final,5p,times,twocolumn]{elsarticle}

\usepackage[colorlinks=true,linkcolor=red,citecolor=blue,urlcolor=blue]{hyperref}
\usepackage{threeparttable}
\usepackage{amssymb}
\usepackage{epsfig} 
\usepackage{amsmath} 
\usepackage{graphicx}
\usepackage{color}
\usepackage{float}
\usepackage{xcolor}
\usepackage{amssymb}
\usepackage{enumitem}
\usepackage{natbib}
\usepackage[T1]{fontenc}

\usepackage{amssymb}

\journal{}

\begin{document}

\begin{frontmatter}



\title{Direct URCA process in light of PREX-2}


\author[1]{Vivek Baruah Thapa}

\affiliation[1]{organization={Indian Institute of Technology Jodhpur},
            city={Jodhpur},
            postcode={342037}, 
            state={Rajasthan},
            country={India}}

\author[1]{Monika Sinha}


\begin{abstract}
We study the implications of the recent development in nuclear symmetry energy constraints from PREX-2 data on dense matter equation of state and its impact on dURCA threshold density.
In this work, we construct the equation of state within the framework of covariant density functional theory implementing coupling schemes of non-linear and density-dependent models and exploring the coupling parameter space of isovector-vector meson to baryons constrained by the isospin asymmetry parameter values deduced from recent PREX-2 data.
The modified parameter sets are applied to evaluate the dense matter properties.
We find that the updated data suggests the occurence of dURCA process within neutron star even with mass as low as one solar mass.
\end{abstract}




\end{frontmatter}


\section{Introduction}\label{sec:intro}
Neutron stars (NSs) provide the most suitable environment to study dense matter behaviours \cite{1996cost.book.....G, weber2017pulsars} as they are highly compact objects containing matter at a few times nuclear saturation density ($n_0$).
The properties of that kind of matter can not be studied from terrestrial experimental data as it is difficult to produce that much high density on any terrestrial facility. Hence one way to understand matter at that density region is by extrapolating nuclear matter properties obtained in nuclear experiments. However, the matter inside NSs are highly asymmetric. Hence the nuclear symmetry energy and its density dependence play an important role to fix the dense matter properties inside NSs.
Earlier, studies \cite{2014EPJA...50...40L, 2015PhRvC..92f4304R, 2017ApJ...848..105T} were accomplished to constrain the values of nuclear symmetry energy $E_{\text{sym}}$ and its slope $L_{\text{sym}}$ at $n_0$ based on data from various astrophysical observations as well as terrestrial experiments.
The values of $E_{\text{sym}}(n_0)$, $L_{\text{sym}}(n_0)$ were estimated to be in the range $[28.5-34.9]$ MeV and $[30.6-86.8]$ MeV \cite{2017RvMP...89a5007O} respectively based on data from various models.
Another nuclear saturation parameter, the curvature of symmetry energy $K_{\text{sym}}$ has been studied in recent years with predicted range values $-111.8 \pm 71.3$ MeV \cite{2017PhRvC..96b1302M}, $-85^{+82}_{-70}$ MeV \cite{2019ApJ...887...48B}, $-102^{+71}_{-72}$ MeV \cite{2020arXiv200203210Z} at $n_0$ based on nuclear and astrophysical observational data which put additional constraint on dense matter equation of state (EOS).
Recently, there is significant advancement in constraining the values of $E_{\rm{sym}}$ and its dependence on density from
the measurement of neutron skin-thickness ($\Delta R_{np}$).
Ref.~\cite{PhysRevLett.126.172502} reported the updated value of $\Delta R_{np}=0.283 \pm 0.071$ fm of $^{208}\text{Pb}$ from the Lead Radius EXperiment-II (PREX-2) with $\sim 1\%$ precision.
Based on this data, Ref.~\cite{PhysRevLett.126.172503} deduced the isospin asymmetry involving parameter values as $L_{\text{sym}}(n_0)\sim 106 \pm 37$ MeV and $E_{\text{sym}}(n_0)\sim 38.1 \pm 4.7$ MeV.

Among many nuclear matter properties $E_{\rm{sym}}$ and its dependence on density play a crucial role to determine the relative abundance of different isotopic spin projections of nucleons.
On the other hand the variation of relative abundances of different particles with matter density is very important for the threshold of direct URCA (dURCA) process.
Inside the NS core, the constituent particles namely nucleons and leptons remain in the degenerate state. Hence, the Fermi momenta of the particles are determined by their respective number density.  If the Fermi momenta of the participating particles in the dURCA process do not satisfy the triangle condition then dURCA process is forbidden. dURCA processes produce neutrinos expeditiously in dense NS matter which leads to rapid NS cooling \cite{1992ApJ...390L..77P, 2004A&A...424..979B, 2004ARA&A..42..169Y, 2005NuPhA.759..373K, 2019FrASS...6...13P}. Interior to star's core, among many other possible processes, dURCA process is several orders of magnitude more efficient that other neutrino emitting processes \cite{PhysRevLett.66.2701}.
NSs are born as the stellar remnant of type-II supernova explosions with average temperature of the order $\sim 10^{11}$ K. After the birth, the core of the NSs cools down comparatively faster than the crust part by neutrino emission. Due to the uncertainties in nuclear physics sector at high matter density regimes, the occurrence of dURCA process in NS interior lies in question. With the matter EOS constructed from so far available nuclear physics data, cooling by dURCA process for stars with masses in the range M$_{\text{typ}}\sim 1.0-1.5$ M$_{\odot}$ is not admissible \cite{RevModPhys.74.1015, 2006A&A...448..327P, 2002PhRvC..66e5803H}.
In this letter we examine the dURCA threshold in light of the newly obtained nuclear symmetry energy and its slope by PREX-2 experiment \cite{PhysRevLett.126.172503}.

The simplest model of NS interior is the matter composed of nucleons and leptons (electrons and muons).
Various approaches both phenomenological \cite{1972PhRvC...5..626V,2001A&A...380..151D,2014PhRvC..89d5807B} as well as microscopic \cite{1998PhRvC..58.1804A,2015RvMP...87.1067C,2019PhRvC.100d5803L} have been explored to understand the dense matter composition in many studies.
Phenomenological Covariant Density Functional (CDF) approach has been widely implemented \cite{Fortin_PRC_2016, Chen_PRC_2007, Zhu_PRC_2016,Sahoo_PRC_2018,Kolomeitsev_NPA_2017, Li_PLB_2018,Li2019ApJ, Ribes_2019,Li2020PhRvD} to study NS matter in which the effective interaction between baryons are accounted via exchange of several mesons.
Within this formalism, the model parameters are decided in such a way that the model can reproduce the nuclear matter properties at $n_0$. CDF models can opportunely describe the finite nuclei properties viz. effective Dirac nucleon mass, 
saturation energy, 
incompressibility, 
symmetry energy, 
its slope 
and curvature 
at $n_0$. 

The interaction through isovector-vector $\rho$ meson is closely related to the iso-spin asymmetric nature of the matter. Hence, phenomenologically the $\rho$-meson coupling with nucleon determines the $E_{\text{sym}}$. The density variation of $E_{\rm{sym}}$ can be implemented in the matter considering the density dependent isovector-vector ($\rho$-meson) coupling with nucleons. Consequently, we implement the density-dependence of ($\rho$-meson) coupling with GM1, GM2, GM3 \cite{1991PhRvL..67.2414G} in non-linear (NL) sector and tune the density dependent $\rho$-meson coupling parameter
space in NL sector as well as with DD1 \cite{2005PhRvC..71f4301T}, DD2 \cite{2010PhRvC..81a5803T},
DD-ME1 \cite{PhysRevC.66.024306}, DD-ME2 \cite{2005PhRvC..71b4312L},
DD-MEX \cite{TANINAH2020135065} coupling parametrizations in density-dependent (DD) sector within CDF model scheme to get the newly obtained values of $E_{\rm{sym}}$ and $L_{\rm{sym}}$. Then we examine the dURCA threshold with this newly obtained matter.


This paper is organized as follows.
The formalism based on CDF model for constructing the EOS is briefly described in sec.-\ref{sec:formalism}.
Nuclear symmetry energy influence on dense matter and subsequent implications on NS observables are displayed and discussed in sec.-\ref{sec:eos}, \ref{sec:star}, \ref{sec:DUrca}.
Finally, the summary and concluding remark of this work are provided in sec.-\ref{sec:summary}.

\textit{Conventions}: We implement the natural units $G=\hbar=c=1$ throughout the work.

\section{CDF Model}\label{sec:formalism}

The CDF model implemented in this work to construct the EOS is briefly discussed in this section.
The Lagrangian density describing the interaction between the mesons ($\sigma$, $\omega$, $\rho$) and nucleons is given by \cite{1996cost.book.....G}
\begin{equation}\label{eqn.01}
\begin{aligned}
\mathcal{L} & = \sum_{N\equiv n,p} \bar{\psi}_N(i\gamma^{\mu} D_{\mu} - m^{*}_N) \psi_N + \frac{1}{2}(\partial_{\mu}\sigma\partial^{\mu}\sigma - m_{\sigma}^2 \sigma^2) \\
 & -  \frac{1}{4}\omega_{\mu\nu}\omega^{\mu\nu} + \frac{1}{2}m_{\omega}^2\omega_{\mu}\omega^{\mu} - \frac{1}{4}\boldsymbol{\rho}_{\mu\nu} \cdot \boldsymbol{\rho}^{\mu\nu} + \frac{1}{2}m_{\rho}^2\boldsymbol{\rho}_{\mu} \cdot \boldsymbol{\rho}^{\mu}  \\
 & - \text{U}(\sigma) + \sum_{l\equiv e^-,\mu^-} \bar{\psi}_l (i\gamma_{\mu} \partial^{\mu} - m_l)\psi_l,
\end{aligned}
\end{equation}
where $\psi_N$, $\psi_l$ represent the Dirac fields of nucleons and leptons respectively, and the covariant derivative
$D_\mu = \partial_\mu + ig_\omega \omega_\mu + ig_\rho \boldsymbol{\tau}_3 \cdot \boldsymbol{\rho}_{\mu}$ with $g_\omega$ and $g_\rho$ being the coupling constnat for coupling of nucleons with $\omega$ and $\rho$ mesons respectively.
The scalar self-interaction term present only in NL CDF models is given by, $\text{U}(\sigma)=(1/3)g_2\sigma^3 + (1/4)g_3\sigma^4$ with $g_2$, $g_3$ being the coefficients of self-interactions.
The re-arrangement term necessary to maintain thermodynamic consistency in case of density-dependent coupling schemes is given by \cite{1997PhRvC..55..540L}
\begin{equation}
\begin{aligned}\label{eqn.02}
	\Sigma^{r} & = \sum_{N} \left[ \frac{\partial g_\omega}{\partial n}\omega_{0}n_{N} - \frac{\partial g_\sigma}{\partial n} \sigma n_{N}^s + \frac{\partial g_\rho}{\partial n} \rho_{03} \boldsymbol{\tau}_3 n_{N} \right].
\end{aligned}
\end{equation}
Here, the vector and scalar number densities are denoted by $n_N=\langle\bar{\psi} \gamma^0 \psi \rangle$ and $n^s_N= \langle\bar{\psi_N} \psi_N \rangle$ respectively and $g_\sigma$ is coupling constant for coupling between nucleons and $\sigma$ mesons.
Charge neutrality and global baryon number conservation constraints are taken into consideration as well while evaluating dense matter EOSs.

The coupling parameters and mass of the mesons are considered in the way that the model can reproduce certain properties of nuclear matter at saturation density. One of the saturation properties the $E_{\text{sym}}$ and its density dependence are important to get the idea about the isovector-vector meson coupling constants $g_{\rho}$ and its variation with density. In terms of energy density we can get the values of $E_{\text{sym}}$ by
the Taylor's expansion of the energy density of symmetric nuclear matter (NM) in terms of neutron-proton asymmetry factor, $\alpha=(n_n-n_p)/n$ 
\begin{equation} \label{eqn.03}
\varepsilon (n,\alpha)=\varepsilon (n,0) + \frac{1}{2} \left[\frac{\partial^2 \varepsilon (n,\alpha)}{\partial \alpha^2} \right]_{\alpha=0} \alpha^2 + \mathcal{O}(\alpha^4),
\end{equation}
where $n$, $n_n$, $n_p$ denote the baryon number, vector number densities of neutron and proton respectively.
The coefficient of second term in eq.-\eqref{eqn.03} refers to the nuclear symmetry energy $E_{\text{sym}}(n)$.
Subsequent expansion of $E_{\text{sym}}(n)$ around $n_0$ provides \cite{MATSUI1981365, PhysRevC.90.044305}
\begin{equation} \label{eqn.04}
\begin{aligned}
E_{\text{sym}}(n) & = E_{\text{sym}}(n_0) + L_{\text{sym}}(n_0) \zeta + \frac{1}{2} K_{\text{sym}}(n_0) \zeta^2 \\ & + \mathcal{O} (\zeta^3),
\end{aligned}
\end{equation}
where $\zeta=(n-n_0)/3n_0$, $E_{\text{sym}}(n_0)$ denotes the nuclear symmetry energy at nuclear saturation density.
The slope and curvature of symmetry energy coefficient at $n_0$ are represented respectively by
\begin{equation} \label{eqn.05}
\begin{aligned}
L_{\text{sym}}(n_0) & = 3 n_0 \left[\frac{\partial E_{\text{sym}} (n)}{\partial n} \right]_{n=n_0}, \\
K_{\text{sym}}(n_0) & = 9 n_0^2 \left[\frac{\partial^2 E_{\text{sym}} (n)}{\partial n^2} \right]_{n=n_0}.
\end{aligned}
\end{equation}
Recently, from PREX-2 data the constraints on the $E_{\text{sym}}$ and $L_{\text{sym}}$ has been updated as discussed in sec.-\ref{sec:intro}. We use this updated data to evaluate the coupling constant parameters and hence the corresponding matter and star properties. As mentioned earlier, the values of $E_{\text{sym}}$ and $L_{\text{sym}}$ in the matter can be tuned by fixing the parameters in density dependent $g_\rho$. We implement density dependent $g_\rho$ in NL sector too and use the identical empirical form of its density dependence \cite{1997PhRvC..55..540L} as
\begin{equation} \label{eqn.06}
g_{\rho}(n)= g_{\rho}(n_{0}) e^{-a_{\rho}(n/n_0-1)}.
\end{equation}
where $a_\rho$ is a constant parameter as used in DD sector.


\begin{figure*}[!htb]
   \begin{minipage}{0.48\textwidth}
     \centering
     \includegraphics[width=8.5cm, keepaspectratio ]{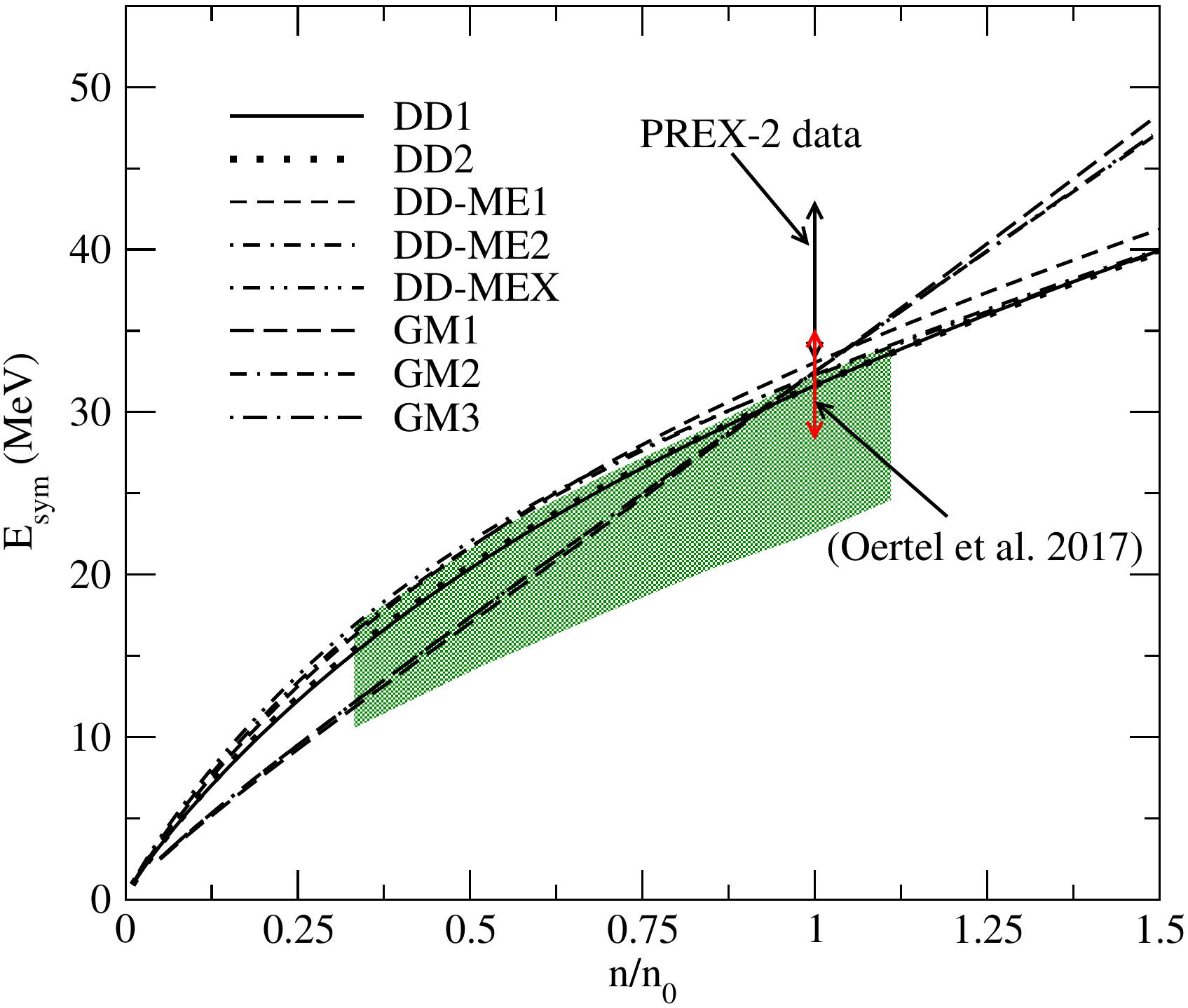}
   \end{minipage}\hfill
   \begin{minipage}{0.48\textwidth}
     \centering
     \includegraphics[width=8.5cm, keepaspectratio ]{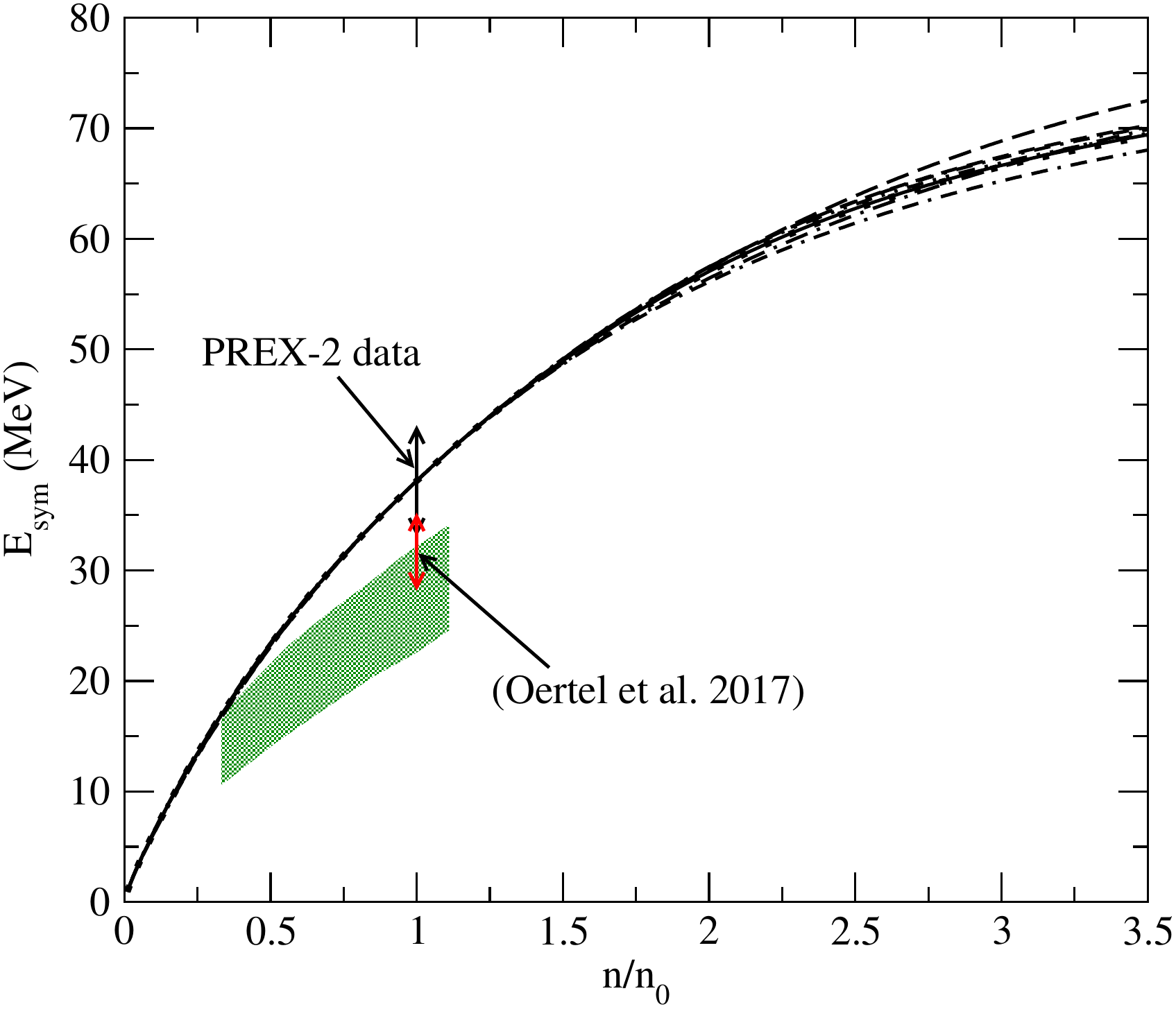}
   \end{minipage}
   \caption{Left panel: $E_{\text{sym}}$ as a function of baryon number density for all the considered parametrization models and right panel: Similar to left panel, but with isovector couplings of the considered parametrizations calibrated to reproduce values of $E_{\text{sym}}(n_0)=38.1$ MeV and $L_{\text{sym}}(n_0)=75$ MeV following PREX-2 data. The shaded region represents the constraint from heavy-ion collisions \cite{2009PhRvL.102l2701T, TSANG2011400}. The red vertical range denote the empirical range of $E_{\text{sym}}(n_0)$ following recent model calculations based on experimental data \citep{2017RvMP...89a5007O}. The black vertical range is obtained from updated PREX-2 data \cite{PhysRevLett.126.172503}. The different curves represent various coupling models as labelled in the figure.}     \label{fig.01}
\end{figure*}

\section{Equation of state parameters}\label{sec:eos}

As discussed above, first we determine the CDF model parametrizations from the newly estimated range of $E_{\text{sym}}$ and $L_{\text{sym}}$. In this purpose, we employ the CDF model with coupling parametrizations of GM1, GM2, GM3 \cite{1991PhRvL..67.2414G} in NL sector and DD1 \cite{2005PhRvC..71f4301T}, DD2 \cite{2010PhRvC..81a5803T}, DD-ME1 \cite{PhysRevC.66.024306}, DD-ME2 \cite{2005PhRvC..71b4312L}, DD-MEX \cite{TANINAH2020135065} in DD sector.

\begin{table*} [h!]
\centering
\caption{The nuclear properties of the considered CDF coupling models at $n_0$. Here $E_0$, $K_0$ denote the binding energy per nucleon and incompressibility respectively.}
\begin{threeparttable}
\scalebox{0.93}{
\begin{tabular}{cccccccc}
\hline \hline
 \multicolumn{2}{c}{CDF} & $n_0$ & $-E_0$ & $K_0$ & $E_{sym}$ & $L_{sym}$ & $K_{sym}$ \\
  \multicolumn{2}{c}{Model} & (fm$^{-3}$) & (MeV) & (MeV) & (MeV) & (MeV) & (MeV) \\
\hline
 & DD1 & 0.1487 & $16.02$ & 240 & 31.60 & 55.95 & $-95.24$ \\
 & DD2 & 0.149 & $16.02$ & 242.7 & 32.73 & 54.97 & $-93.24$ \\
DD & DD-ME1 & 0.152 & $16.20$ & 244.5 & 33.10 & 55.37 & $-101.07$ \\
 & DD-ME2 & 0.152 & $16.14$ & 250.89 & 32.30 & 51.25 & $-87.31$ \\
& DD-MEX & 0.1518 & $16.14$ & 267.06 & 32.27 & 49.58 & $-71.47$ \\
 \hline
 & GM1 & 0.153 & $16.30$ & 300 & 32.50 & 93.86 & 17.91 \\
NL & GM2 & 0.153 & $16.30$ & 300 & 32.50 & 89.29 & $-11.98$ \\
 & GM3 & 0.153 & $16.30$ & 240 & 32.50 & 89.63 & $-6.46$ \\
\hline
\end{tabular}
}
\end{threeparttable}
\label{tab:01}
\end{table*}
\begin{figure} [h!]
  \begin{center}
\includegraphics[width=8.5cm,keepaspectratio ]{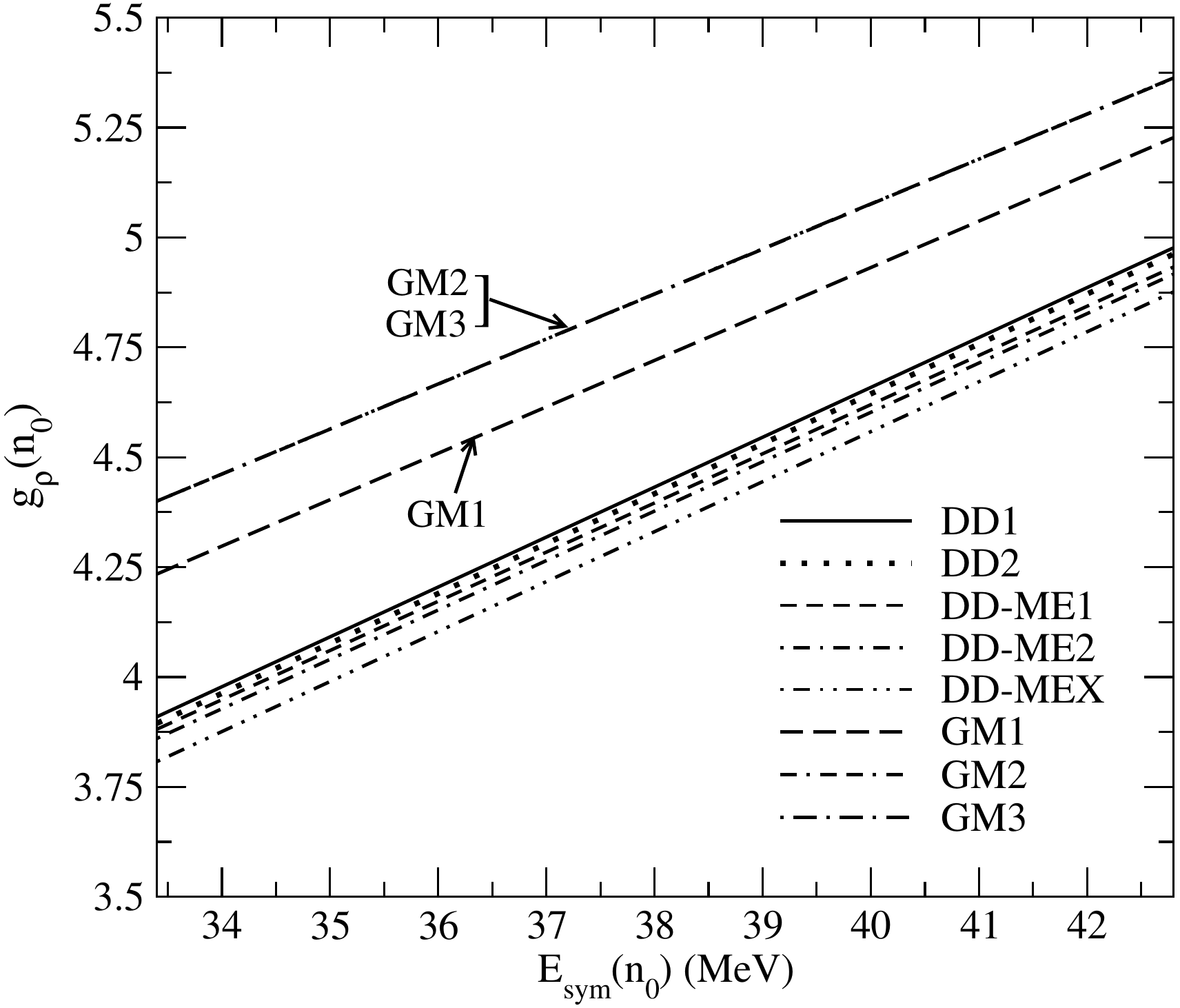}
\caption{Variation in the values of the isovector coupling parameter to baryons ($g_{\rho}$) at $n_0$ with $E_{\text{sym}}(n_0)$. The various curves represent the same coupling models as labelled in fig.-\ref{fig.01}.} 
\label{fig.07}
\end{center}
\end{figure}
\begin{table*} 
\centering
\caption{Calibrated isovector-vector $\rho$ meson couplings to baryons at $n_0$ corresponding to $E_{\text{sym}} (n_0)=38.1$ MeV for different parametrizations considered in this work.}
\begin{threeparttable}
\scalebox{0.93}{
\begin{tabular}{ccccccccc}
\hline \hline
 CDF Model & DD1 & DD2 & DD-ME1 & DD-ME2 & DD-MEX & GM1 & GM2 & GM3 \\
\hline
 $g_{\rho}(n_0)$ & 4.45450 & 4.44010 & 4.41853 & 4.40010 & 4.35400 & 4.74027 & 4.88940 & 4.89040 \\
 \hline
\end{tabular}
}
\end{threeparttable}
\label{tab:02}
\end{table*}
With so far known ranges of the nuclear saturation properties, the above prametrizations are fixed to reproduce the same as shown in table-\ref{tab:01}.
With these existing parametrizations, the density-dependence of $E_{\text{sym}}(n)$ is shown in the left panel of fig.-\ref{fig.01}.
It can be seen that the mentioned coupling models do not satisfy the recent constraint of $E_{\text{sym}}(n_0)$ deduced from PREX-2 data as also evident from table-\ref{tab:01}.
To get the updated ranges of $E_{\text{{sym}}}$, we tune the values of $g_\rho(n_0)$ and $a_\rho$ for different parametrizations in both NL and DD schemes keepeing their other parameters same.
The corresponding tuned parametrizations to produce the isospin asymmetry parameters, $E_{\text{sym}}=38.1$ MeV and $L_{\text{sym}}=75$ MeV at $n_0$ fulfilling PREX-2 data are displayed in the right panel of fig.-\ref{fig.01}.
Since both the parameter values are taken to be identical for all the coupling models, the curves for different EOSs nearly overlap upto $\sim 1.5~n_0$. Beyond that density range, the curves deviate. This is because of the difference in density-dependent isovector coupling behavior at high matter density regimes. It is observed that with higher $E_{\text{sym}}(n_0)$ values, the constraint from heavy-ion collisions is satisfied only at sub-saturation densities $\sim 0.37~n_0$.

\begin{figure} [h!]
  \begin{center}
\includegraphics[width=8.5cm,keepaspectratio ]{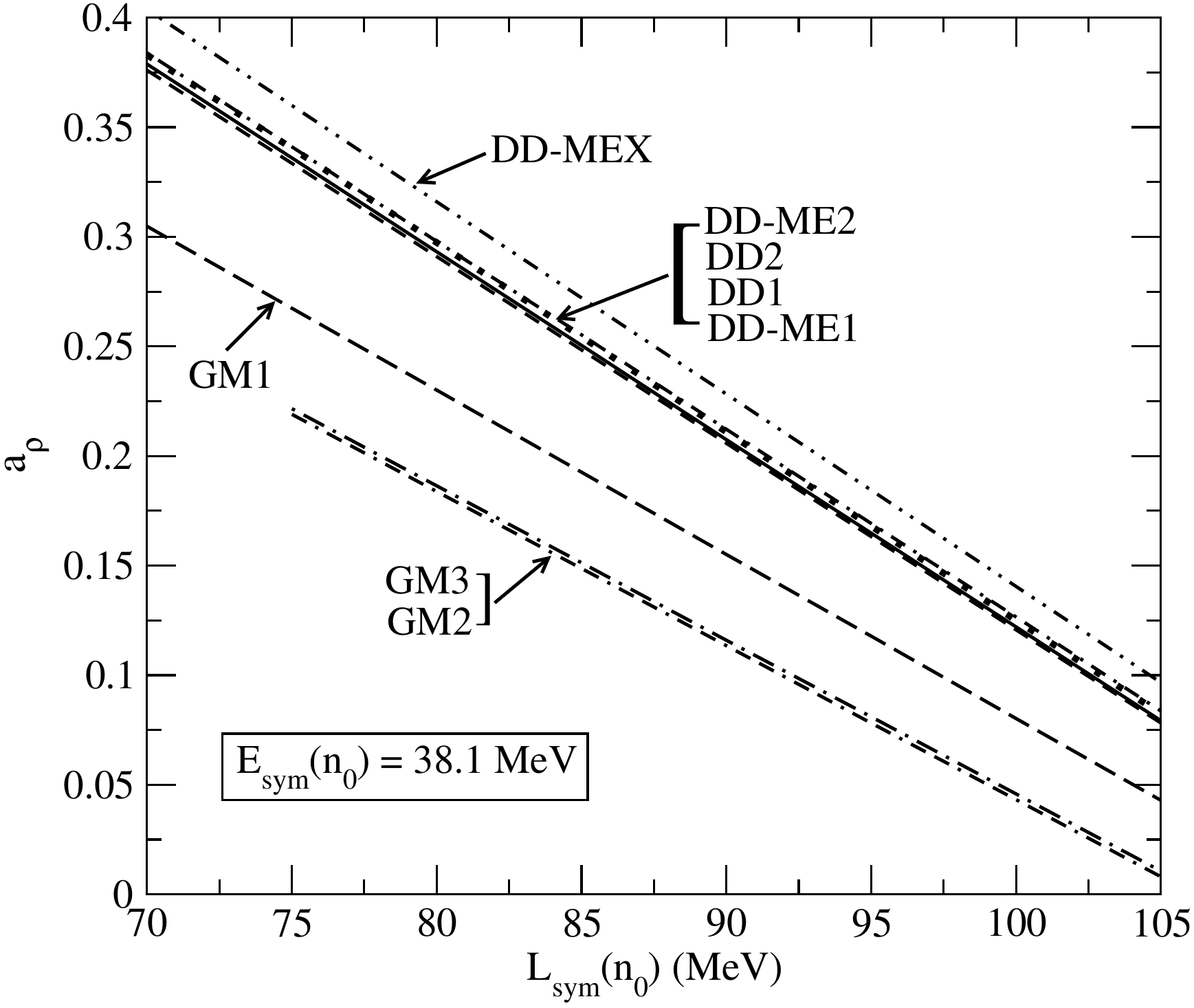}
\caption{Variation in the values of the parameter $a_\rho$ with $L_{\text{sym}}(n_0)$
The various curves represent the same coupling models as labelled in fig.-\ref{fig.01}.}
\label{fig.05}
\end{center}
\end{figure}

For the recently obtained range of $E_{\text{{sym}}}$, the variation of $g_\rho(n_0)$ for different parametrizations is shown in fig.-\ref{fig.07}.
The variation of $E_{\text{{sym}}}$ with density and hence the $L_{\text{{sym}}}$ depends on the choice of the parameter value of $a_\rho$ {\it i.e.} on the density dependence of $g_\rho$. For the recently obtained experimental range of $L_{\text{{sym}}}$, the variation of $a_\rho$ is shown in the fig.-\ref{fig.05} without compromising other nuclear saturation properties. For this evaluation, the value of $E_{\text{{sym}}}(n_0)$ is fixed to $38.1$ MeV inferred from recent PREX-2 data and the corresponding values of $g_\rho(n_0)$ for different parametrization schemes are given in table-\ref{tab:02}.
A strong linear correlation is observed between these two parameters with slope steeper in case of DD models compared to NL ones.

\begin{figure} [h!]
  \begin{center}
\includegraphics[width=8.5cm,keepaspectratio ]{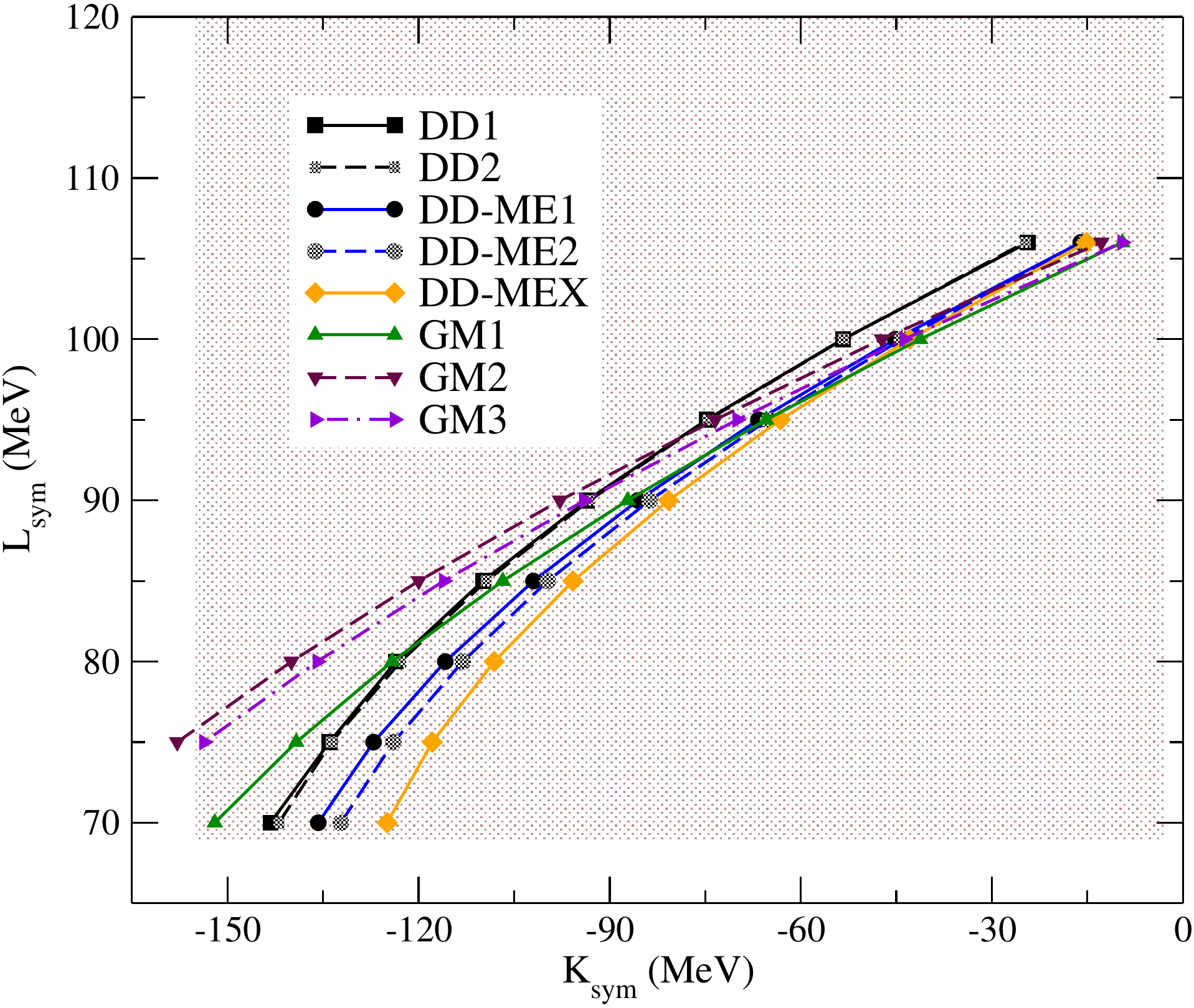}
\caption{Slope and curvature of $E_{\text{sym}}$ for the considered coupling parametrizations. The shaded region represents the empirical range in $L_{\text{sym}}-K_{\text{sym}}$ plane based on terrestrial and astrophysical data \cite{PhysRevLett.126.172503, 2019ApJ...887...48B}.}
\label{fig.02}
\end{center}
\end{figure}

\begin{figure*} [t!]
  \begin{center}
\includegraphics[width=14.0cm,keepaspectratio ]{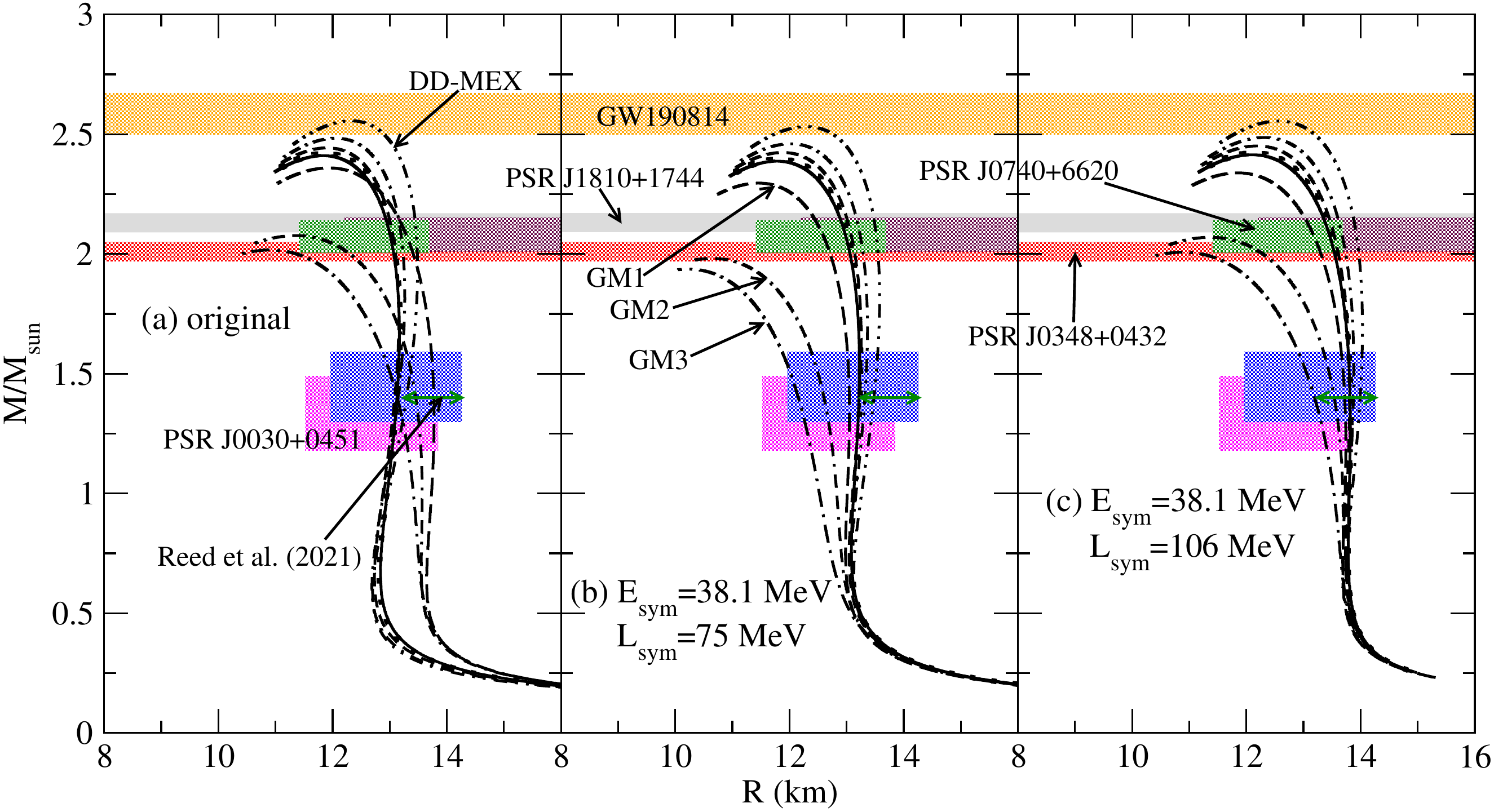}
\caption{TOV solutions corresponding to the EOSs evaluated from left panel: original (no adjustment to $\rho$-meson coupling), middle panel: adjusted to $E_{\text{sym}}=38.1$ MeV, $L_{\text{sym}}=75$ MeV at $n_0$ and right panel: adjusted to $E_{\text{sym}}=38.1$ MeV, $L_{\text{sym}}=106$ MeV at $n_0$. The astrophysical observable constraints from GW190814 (secondary component) \cite{2020ApJ...896L..44A}, PSR J$0740+6620$ \cite{2021arXiv210506980R, 2021arXiv210506979M}, PSR J$1810+1744$ \cite{2021ApJ...908L..46R}, PSR J$0348+0432$ \cite{2013Sci...340..448A} and PSR J$0030+0451$ \cite{2019ApJ...887L..24M, 2019ApJ...887L..21R} are represented by shaded regions. The horizontal line represent the joint radius constraint on a canonical NS deduced from PREX-2 and astrophysical data \cite{PhysRevLett.126.172503}.}
\label{fig.03}
\end{center}
\end{figure*}

We verify the new parametrization with the constrained valued of $K_{\rm{sym}}(n_0)$ as  obtained from empirical parametrization based on Monte Carlo simulation approach by fitting the spectra of quiescent low-mass X-ray binaries \cite{2019ApJ...887...48B}.
Fig.-\ref{fig.02} provides the variation of $L_{\text{sym}}$ with its corresponding curvature $K_{\text{{sym}}}$ at $n_0$.
It can be observed that the considered coupling models fall well within the overlapped constrained region.
Along lower end values of $L_{\text{sym}}(n_0)$, the curvature of $E_{\text{sym}}$ values diverge, while at the higher end of $L_{\text{sym}}(n_0)$, they converge.
Based on $-155 \leqslant K_{\text{sym}}(n_0)/\text{MeV} \leqslant -3$ constraint incorporated here, it can be seen that the upper limit of $L_{\text{sym}}(n_0)$ tends to be around $\sim 110$ MeV.

\section{Star structure}\label{sec:star}

Getting the ranges of parameters to fix the EOS of matter, we study the properties of stars composed of that kind of matter. We obtain the mass-radius structure of the stars with the Baym-Pethick-Sutherland (BPS) \cite{1971ApJ...170..299B} + Baym-Bethe-Pethick (BBP) \cite{1971NuPhA.175..225B} EOS for the crust for non-rotating, spherically symmetric configurations. For different EOSs we show the mass-radius relation in fig.-\ref{fig.03}. The left panel shows the M-R relation with existing parametrizations. The middlle and right panels show different parametrizations of $g_\rho$ producing different values of $E_{\text{{sym}}}$ and $L_{\text{{sym}}}$ corresponding to the their newly obtained range from PREX-2. It is observed that with higher values of $L_{\rm{sym}}$, the radius of the intermediate mass star increases. This evantually restricts the upper limit of $L_{\rm{sym}}$ from NICER measurment of radius of intermediate mass stars. It is to be noted that with lower values of $L_{\rm{sym}}= 75$, the NL coupling models GM2, GM3 fail to satisfy the observed minimum value of maximum attainable mass of the NSs.
This tallies with the results from Refs.-\cite{1996NuPhA.606..508M, 2001PhRvL..86.5647H} which reports the small affect of $L_{\rm{sym}}$ on maximum mass configurations.

\begin{table*}
\centering
\caption{Threshold estimates for nucleonic DU process in NS matter. $x_{\text{DU}}$, $n_{\text{DU}}$, $M_{\text{DU}}$ represent the minimum proton fraction for DU onset, corresponding matter density and threshold masses of the NSs respectively. $E_{\text{sym}}$ at $n_0$ is fixed to 38.1 MeV. Here, ``\textit{OL}" denote the respective original coupling parameterizations.}
\begin{threeparttable}
\scalebox{0.75}{
\begin{tabular}{cccc|ccc|ccc|cccccc}
\hline \hline
 \multicolumn{16}{c}{(i) Density-dependent models} \\
 \hline
 & \multicolumn{3}{|c}{DD1} & \multicolumn{3}{|c}{DD2} & \multicolumn{3}{|c}{DD-ME1} & \multicolumn{3}{|c}{DD-ME2} & \multicolumn{3}{|c}{DD-MEX} \\
 \cline{2-16}
 & \multicolumn{1}{|c}{} & \multicolumn{2}{c|}{$L_{\text{sym}}$} & & \multicolumn{2}{c|}{$L_{\text{sym}}$} & & \multicolumn{2}{c|}{$L_{\text{sym}}$} & & \multicolumn{2}{c|}{$L_{\text{sym}}$} & & \multicolumn{2}{c}{$L_{\text{sym}}$} \\
 & \multicolumn{1}{|c}{\textit{OL}} & 75 & 106 & \textit{OL} & 75 & 106 & \textit{OL} & 75 & 106 & \textit{OL} & 75 & 106 & \multicolumn{1}{|c}{\textit{OL}} & 75 & 106 \\
\hline
\multicolumn{1}{c|}{$x_{\text{DU}}$} & $-$ & 0.13586 & 0.13022 & $-$ & 0.13598 & 0.13024 & $-$ & 0.13554 & 0.13051 & $-$ & 0.13548 & 0.13051 & \multicolumn{1}{|c}{$-$} & 0.13548 & 0.13052 \\
\multicolumn{1}{c|}{$n_{\text{DU}}/n_0$} & $-$ & 2.81 & 1.47 & $-$ & 2.85 & 1.47 & $-$ & 2.63 & 1.48 & $-$ & 2.61 & 1.48 & \multicolumn{1}{|c}{$-$} & 2.61 & 1.48 \\
\multicolumn{1}{c|}{$M_{\text{DU}}/M_\odot$} & $-$ & 1.725 & 0.736 & $-$ & 1.772 & 0.740 & $-$ & 1.684 & 0.762 & $-$ & 1.739 & 0.779 & \multicolumn{1}{|c}{$-$} & 1.878 & 0.809 \\
\hline \hline
 \multicolumn{16}{c}{(i) Non-linear models} \\
 \cline{4-13}
 & \multicolumn{3}{c}{} & \multicolumn{3}{|c}{GM1} & \multicolumn{3}{|c}{GM2} & \multicolumn{3}{|c}{GM3} & \multicolumn{3}{c}{} \\
 \cline{4-13}
 & & & & & \multicolumn{2}{c|}{$L_{\text{sym}}$} & & \multicolumn{2}{c|}{$L_{\text{sym}}$} & & \multicolumn{2}{c}{$L_{\text{sym}}$} & & \multicolumn{2}{c}{} \\
& & & & \textit{OL} & 75 & 106 & \textit{OL} & 75 & 106 & \textit{OL} & 75 & 106 &   \multicolumn{3}{c}{} \\
\cline{4-13}
& & & $x_{\text{DU}}$ & 0.13257 & 0.13575 & 0.13068 & 0.13319 & $-$ & 0.13076 & 0.13311 & $-$ & 0.13065 & \multicolumn{3}{c}{} \\
& & & $n_{\text{DU}}/n_0$ & 1.82 & 2.69 & 1.49 & 1.95 & $-$ & 1.50 & 1.93 & $-$ & 1.49 & \multicolumn{3}{c}{}  \\
& & & $M_{\text{DU}}/M_\odot$ & 1.093 & 1.553 & 0.779 & 1.080 & $-$ & 0.772 & 0.966 & $-$ & 0.712 & \multicolumn{3}{c}{} \\
\cline{4-13}
\end{tabular}
}
\end{threeparttable}
\label{tab:03}
\end{table*}

\begin{figure*} 
  \begin{center}
\includegraphics[width=13.5cm,keepaspectratio ]{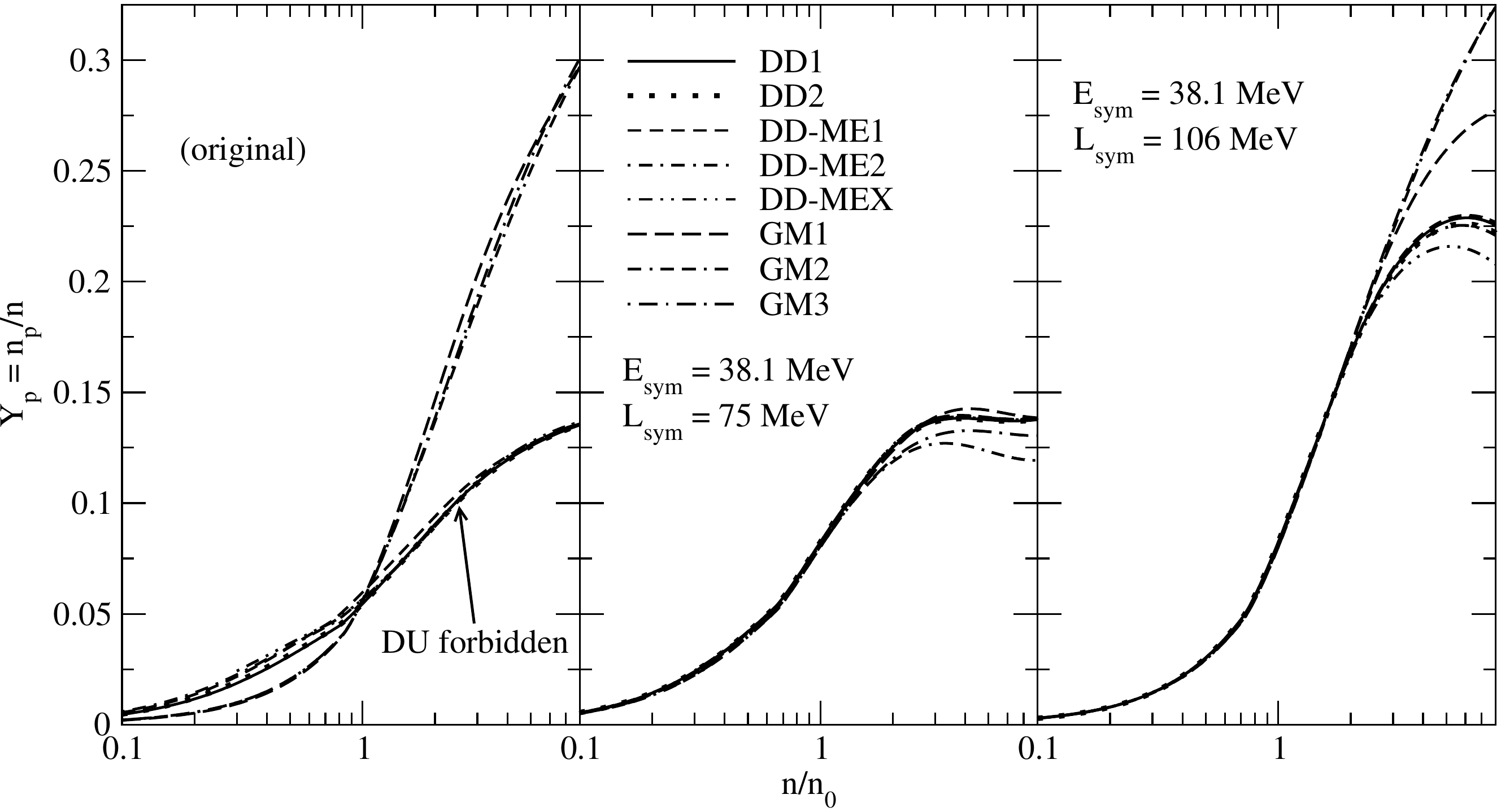}
\caption{The proton fraction (Y$_{\text{p}}$) as a function of baryon number density for various EoSs investigated in this work corresponding to different cases of $\rho$-meson coupling adjustment as in fig.-\ref{fig.03}. The various curves represent the same coupling models as labelled in fig.-\ref{fig.01}.}
\label{fig.04}
\end{center}
\end{figure*}

\begin{figure} [h!]
  \begin{center}
  \includegraphics[width=8.5cm,keepaspectratio ]{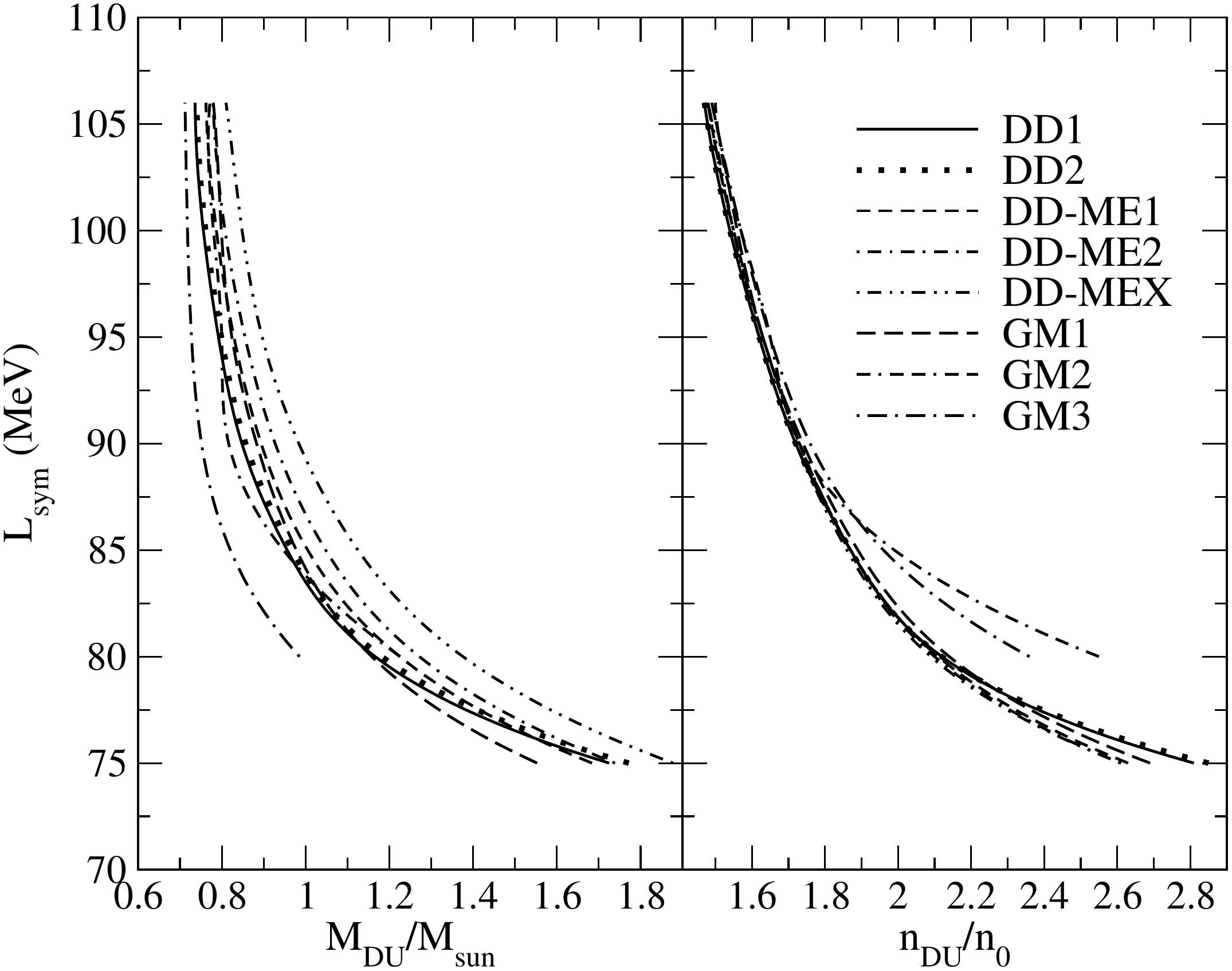}
\caption{Right panel: Threshold densities for nucleonic DU processes and left panel: corresponding threshold NS masses as a function of $L_{\text{sym}}(n_0)$ for different coupling models considered in this work. The various curves represent the same parametrizations as in fig-\ref{fig.01}.}
\label{fig.06}
\end{center}
\end{figure}

\section{Direct URCA process in Neutron stars}\label{sec:DUrca}

Now we study relative abundances of constituent particles with newly obtained EOS parametrizations of $g_\rho$ and consequently the possibility of appearance of dURCA process - the most efficient neutrino emission process. The dURCA process is given by
\begin{equation}\label{eqn.07}
n \rightarrow p + e^- + \bar{\nu}_e, \quad p + e^- \rightarrow n + \nu_e.
\end{equation}
This process activates only when the proton fraction surpasses the critical value, $x_{\text{DU}}$ following the inequality, $p_{F_p}+p_{F_e} \geqslant p_{F_n}$ where $p_{F_p}$, $p_{F_n}$, $p_{F_e}$ represent the Fermi momenta of neutron, proton and electron respectively.
The critical value of proton fraction for dURCA threshold is estimated considering the charge neutrality condition and given as \cite{PhysRevC.74.035802}, $x_{\text{DU}}=\left[1+(1+x_e^{1/3})^3 \right]^{-1}$ with $x_e=n_e/(n_e+n_\mu)$ denoting the leptonic fraction. 

As mentioned the relative abundance of protons depends on the behaviour of the $E_{\rm{sym}}$. Hence, with new parametrizations the proton fraction differs from the existing parametrizations. The variation of the relative abundance of protons with baryon number density is shown in the fig.-\ref{fig.04} for different EOSs. The left panel shows the proton fraction for the existing coupling parametrizations. However, the newly obtained ranges of $E_{\rm{sym}}$ and $L_{\text{sym}}$ changes proton fractions and hence the threshold of dURCA process.
It is observed that in EoSs with higher $L_{\text{sym}}(n_0)$ values, the proton fraction exceeds $x_{\text{DU}}$ at comparatively lower densities. The values of $x_{\text{DU}}$ and corresponding $n_{\text{DU}}$ is given in table-\ref{tab:03}.
In case of $L_{\text{sym}}(n_0)$ values corresponding to the original coupling parametrizations, it is observed that no existing coupling model in DD sector allows dURCA to come into picture, while in case of NL sector, threshold of dURCA process occurs at around $\sim 2~n_0$ even with existing parametrizations. With the newly obtained value of $L_{\text{sym}}(n_0)=106$ MeV, all the parmetrizations allow dURCA process with lower $n_{\text{DU}} \sim 1.5 n_0$. This can be clearly understood from the right panel as proton fraction is almost identical for different EoSs.  However, in the lower admissible range if we consider $L_{\text{sym}}(n_0)=75$ MeV dURCA becomes forbidden in case of NL sector except GM1 parametrization. The stars with central densities less than $n_{\text{DU}}$ will not experience dURCA in any portion of the star. The mass of stars with central density equal to respective $n_{\text{DU}}$ for different EOSs are also tabulated in table-\ref{tab:03} as $M_{\text{DU}}$. The variation of $n_{\text{DU}}$ and $M_{\text{DU}}$ in the newly obtained range of $L_{\rm{sym}}$ is shown in two panels of fig.-\ref{fig.06}.
All the coupling models considered in this work show similar dURCA onset behavior when varied with slope of symmetry energy parameter.

\section{Summary and Conclusions}\label{sec:summary}

We have constructed the matter EOS within the CDF formalism with NL and DD scheme to produce the newly obtained values of $E_{\rm{sym}}$ and $L_{\rm{sym}}$. To evaluate that, we have tuned the parametrization related to the nucleon-isovector meson $\rho$ coupling. We have also studied the dURCA thershold with this new matter. The dURCA process is observed to be immensely influenced by the density dependence of symmetry energy. This is because of the fact that the latter determines the proton fraction in dense matter.
This relates with the results from Refs.-\cite{PhysRevC.84.065810, 2014EPJA...50...44P}.
Larger values of $L_{\text{sym}}$ at $n_0$ disfavors higher neutron-proton asymmetry, consequently favoring dURCA process at early matter densities.  From astrophysical observations, lower $L_{\rm{sym}}$ values are discarded. Consequently, with all admissible parametrizations discussed in this work the dURCA is allowed of stars with mass ranging from $1 M_\odot$.
Refined results from PREX and upcoming Calcium Radius EXperiment (CREX) \cite{CREX2021a, CREX2021b} may provide further insight in this regard.

Hence, dURCA is possible even in case of intermediate and low mass star. With this conclusion, the cooling rate of the NSs should be revisited as dURCA is the most effective process in neutrino cooling.
This is beyond the scope of this study and will be addressed in future works.




 \bibliographystyle{elsarticle-num} 
 \bibliography{references}

\begin{thebibliography}{10}
\expandafter\ifx\csname url\endcsname\relax
  \def\url#1{\texttt{#1}}\fi
\expandafter\ifx\csname urlprefix\endcsname\relax\def\urlprefix{URL }\fi
\expandafter\ifx\csname href\endcsname\relax
  \def\href#1#2{#2} \def\path#1{#1}\fi

\bibitem{1996cost.book.....G}
N.~K. {Glendenning}, {Compact Stars, (Springer-Verlag, New York, 2007), 2nd
  ed.}, 1996.

\bibitem{weber2017pulsars}
F.~Weber, \href{https://books.google.co.in/books?id=SSw2DwAAQBAJ}{Pulsars as
  Astrophysical Laboratories for Nuclear and Particle Physics}, Series in High
  Energy Physics, Cosmology and Gravitation, CRC Press, 2017.
\newline\urlprefix\url{https://books.google.co.in/books?id=SSw2DwAAQBAJ}

\bibitem{2014EPJA...50...40L}
J.~M. {Lattimer}, A.~W. {Steiner}, {Constraints on the symmetry energy using
  the mass-radius relation of neutron stars}, European Physical Journal A 50
  (2014) 40.
\newblock \href {http://arxiv.org/abs/1403.1186} {\path{arXiv:1403.1186}},
  \href {https://doi.org/10.1140/epja/i2014-14040-y}
  {\path{doi:10.1140/epja/i2014-14040-y}}.

\bibitem{2015PhRvC..92f4304R}
X.~{Roca-Maza}, X.~{Vi{\~n}as}, M.~{Centelles}, B.~K. {Agrawal}, G.~{Col{\`o}},
  N.~{Paar}, J.~{Piekarewicz}, D.~{Vretenar}, {Neutron skin thickness from the
  measured electric dipole polarizability in $^{68}$Ni$^{120}$Sn and
  $^{208}$Pb}, \prc 92~(6) (2015) 064304.
\newblock \href {http://arxiv.org/abs/1510.01874} {\path{arXiv:1510.01874}},
  \href {https://doi.org/10.1103/PhysRevC.92.064304}
  {\path{doi:10.1103/PhysRevC.92.064304}}.

\bibitem{2017ApJ...848..105T}
I.~{Tews}, J.~M. {Lattimer}, A.~{Ohnishi}, E.~E. {Kolomeitsev}, {Symmetry
  Parameter Constraints from a Lower Bound on Neutron-matter Energy}, \apj
  848~(2) (2017) 105.
\newblock \href {http://arxiv.org/abs/1611.07133} {\path{arXiv:1611.07133}},
  \href {https://doi.org/10.3847/1538-4357/aa8db9}
  {\path{doi:10.3847/1538-4357/aa8db9}}.

\bibitem{2017RvMP...89a5007O}
M.~{Oertel}, M.~{Hempel}, T.~{Kl{\"a}hn}, S.~{Typel}, {Equations of state for
  supernovae and compact stars}, Reviews of Modern Physics 89~(1) (2017)
  015007.
\newblock \href {http://arxiv.org/abs/1610.03361} {\path{arXiv:1610.03361}},
  \href {https://doi.org/10.1103/RevModPhys.89.015007}
  {\path{doi:10.1103/RevModPhys.89.015007}}.

\bibitem{2017PhRvC..96b1302M}
C.~{Mondal}, B.~K. {Agrawal}, J.~N. {De}, S.~K. {Samaddar}, M.~{Centelles},
  X.~{Vi{\~n}as}, {Interdependence of different symmetry energy elements}, \prc
  96~(2) (2017) 021302.
\newblock \href {http://arxiv.org/abs/1708.03846} {\path{arXiv:1708.03846}},
  \href {https://doi.org/10.1103/PhysRevC.96.021302}
  {\path{doi:10.1103/PhysRevC.96.021302}}.

\bibitem{2019ApJ...887...48B}
N.~{Baillot d'Etivaux}, S.~{Guillot}, J.~{Margueron}, N.~{Webb}, M.~{Catelan},
  A.~{Reisenegger}, {New Constraints on the Nuclear Equation of State from the
  Thermal Emission of Neutron Stars in Quiescent Low-mass X-Ray Binaries}, \apj
  887~(1) (2019) 48.
\newblock \href {http://arxiv.org/abs/1905.01081} {\path{arXiv:1905.01081}},
  \href {https://doi.org/10.3847/1538-4357/ab4f6c}
  {\path{doi:10.3847/1538-4357/ab4f6c}}.

\bibitem{2020arXiv200203210Z}
J.~{Zimmerman}, Z.~{Carson}, K.~{Schumacher}, A.~W. {Steiner}, K.~{Yagi},
  {Measuring Nuclear Matter Parameters with NICER and LIGO/Virgo}, arXiv
  e-prints (2020) arXiv:2002.03210\href {http://arxiv.org/abs/2002.03210}
  {\path{arXiv:2002.03210}}.

\bibitem{PhysRevLett.126.172502}
D.~Adhikari, H.~Albataineh, D.~Androic, K.~Aniol, D.~S. Armstrong, T.~Averett,
  et~al.,
  \href{https://link.aps.org/doi/10.1103/PhysRevLett.126.172502}{Accurate
  determination of the neutron skin thickness of $^{208}\mathrm{Pb}$ through
  parity-violation in electron scattering}, Phys. Rev. Lett. 126 (2021) 172502.
\newblock \href {https://doi.org/10.1103/PhysRevLett.126.172502}
  {\path{doi:10.1103/PhysRevLett.126.172502}}.
\newline\urlprefix\url{https://link.aps.org/doi/10.1103/PhysRevLett.126.172502}

\bibitem{PhysRevLett.126.172503}
B.~T. Reed, F.~J. Fattoyev, C.~J. Horowitz, J.~Piekarewicz,
  \href{https://link.aps.org/doi/10.1103/PhysRevLett.126.172503}{Implications
  of prex-2 on the equation of state of neutron-rich matter}, Phys. Rev. Lett.
  126 (2021) 172503.
\newblock \href {https://doi.org/10.1103/PhysRevLett.126.172503}
  {\path{doi:10.1103/PhysRevLett.126.172503}}.
\newline\urlprefix\url{https://link.aps.org/doi/10.1103/PhysRevLett.126.172503}

\bibitem{1992ApJ...390L..77P}
M.~{Prakash}, M.~{Prakash}, J.~M. {Lattimer}, C.~J. {Pethick}, {Rapid Cooling
  of Neutron Stars by Hyperons and Delta Isobars}, \apjl 390 (1992) L77.
\newblock \href {https://doi.org/10.1086/186376} {\path{doi:10.1086/186376}}.

\bibitem{2004A&A...424..979B}
D.~{Blaschke}, H.~{Grigorian}, D.~N. {Voskresensky}, {Cooling of neutron stars.
  Hadronic model}, \aap 424 (2004) 979--992.
\newblock \href {http://arxiv.org/abs/astro-ph/0403170}
  {\path{arXiv:astro-ph/0403170}}, \href
  {https://doi.org/10.1051/0004-6361:20040404}
  {\path{doi:10.1051/0004-6361:20040404}}.

\bibitem{2004ARA&A..42..169Y}
D.~G. {Yakovlev}, C.~J. {Pethick}, {Neutron Star Cooling}, \araa 42~(1) (2004)
  169--210.
\newblock \href {http://arxiv.org/abs/astro-ph/0402143}
  {\path{arXiv:astro-ph/0402143}}, \href
  {https://doi.org/10.1146/annurev.astro.42.053102.134013}
  {\path{doi:10.1146/annurev.astro.42.053102.134013}}.

\bibitem{2005NuPhA.759..373K}
E.~E. {Kolomeitsev}, D.~N. {Voskresensky}, {Relativistic mean-field models with
  effective hadron masses and coupling constants, and
  {\ensuremath{\rho}}<SUP></SUP> condensation}, \nphysa 759~(3-4) (2005)
  373--413.
\newblock \href {http://arxiv.org/abs/nucl-th/0410063}
  {\path{arXiv:nucl-th/0410063}}, \href
  {https://doi.org/10.1016/j.nuclphysa.2005.05.154}
  {\path{doi:10.1016/j.nuclphysa.2005.05.154}}.

\bibitem{2019FrASS...6...13P}
C.~{Provid{\^e}ncia}, M.~{Fortin}, H.~{Pais}, A.~{Rabhi}, {Hyperonic stars and
  the symmetry energy}, Frontiers in Astronomy and Space Sciences 6 (2019) 13.
\newblock \href {http://arxiv.org/abs/1811.00786} {\path{arXiv:1811.00786}},
  \href {https://doi.org/10.3389/fspas.2019.00013}
  {\path{doi:10.3389/fspas.2019.00013}}.

\bibitem{PhysRevLett.66.2701}
J.~M. Lattimer, C.~J. Pethick, M.~Prakash, P.~Haensel,
  \href{https://link.aps.org/doi/10.1103/PhysRevLett.66.2701}{Direct urca
  process in neutron stars}, Phys. Rev. Lett. 66 (1991) 2701--2704.
\newblock \href {https://doi.org/10.1103/PhysRevLett.66.2701}
  {\path{doi:10.1103/PhysRevLett.66.2701}}.
\newline\urlprefix\url{https://link.aps.org/doi/10.1103/PhysRevLett.66.2701}

\bibitem{RevModPhys.74.1015}
S.~E. Woosley, A.~Heger, T.~A. Weaver,
  \href{https://link.aps.org/doi/10.1103/RevModPhys.74.1015}{The evolution and
  explosion of massive stars}, Rev. Mod. Phys. 74 (2002) 1015--1071.
\newblock \href {https://doi.org/10.1103/RevModPhys.74.1015}
  {\path{doi:10.1103/RevModPhys.74.1015}}.
\newline\urlprefix\url{https://link.aps.org/doi/10.1103/RevModPhys.74.1015}

\bibitem{2006A&A...448..327P}
S.~{Popov}, H.~{Grigorian}, R.~{Turolla}, D.~{Blaschke}, {Population synthesis
  as a probe of neutron star thermal evolution}, \aap 448~(1) (2006) 327--334.
\newblock \href {http://arxiv.org/abs/astro-ph/0411618}
  {\path{arXiv:astro-ph/0411618}}, \href
  {https://doi.org/10.1051/0004-6361:20042412}
  {\path{doi:10.1051/0004-6361:20042412}}.

\bibitem{2002PhRvC..66e5803H}
C.~J. {Horowitz}, J.~{Piekarewicz}, {Constraining URCA cooling of neutron stars
  from the neutron radius of $^{208}$Pb}, \prc 66~(5) (2002) 055803.
\newblock \href {http://arxiv.org/abs/nucl-th/0207067}
  {\path{arXiv:nucl-th/0207067}}, \href
  {https://doi.org/10.1103/PhysRevC.66.055803}
  {\path{doi:10.1103/PhysRevC.66.055803}}.

\bibitem{1972PhRvC...5..626V}
D.~{Vautherin}, D.~M. {Brink}, {Hartree-Fock Calculations with Skyrme's
  Interaction. I. Spherical Nuclei}, \prc 5~(3) (1972) 626--647.
\newblock \href {https://doi.org/10.1103/PhysRevC.5.626}
  {\path{doi:10.1103/PhysRevC.5.626}}.

\bibitem{2001A&A...380..151D}
F.~{Douchin}, P.~{Haensel}, {A unified equation of state of dense matter and
  neutron star structure}, \aap 380 (2001) 151--167.
\newblock \href {http://arxiv.org/abs/astro-ph/0111092}
  {\path{arXiv:astro-ph/0111092}}, \href
  {https://doi.org/10.1051/0004-6361:20011402}
  {\path{doi:10.1051/0004-6361:20011402}}.

\bibitem{2014PhRvC..89d5807B}
S.~S. {Bao}, H.~{Shen}, {Influence of the symmetry energy on nuclear ``pasta''
  in neutron star crusts}, \prc 89~(4) (2014) 045807.
\newblock \href {http://arxiv.org/abs/1405.3837} {\path{arXiv:1405.3837}},
  \href {https://doi.org/10.1103/PhysRevC.89.045807}
  {\path{doi:10.1103/PhysRevC.89.045807}}.

\bibitem{1998PhRvC..58.1804A}
A.~{Akmal}, V.~R. {Pandharipande}, D.~G. {Ravenhall}, {Equation of state of
  nucleon matter and neutron star structure}, \prc 58~(3) (1998) 1804--1828.
\newblock \href {http://arxiv.org/abs/nucl-th/9804027}
  {\path{arXiv:nucl-th/9804027}}, \href
  {https://doi.org/10.1103/PhysRevC.58.1804}
  {\path{doi:10.1103/PhysRevC.58.1804}}.

\bibitem{2015RvMP...87.1067C}
J.~{Carlson}, S.~{Gandolfi}, F.~{Pederiva}, S.~C. {Pieper}, R.~{Schiavilla},
  K.~E. {Schmidt}, R.~B. {Wiringa}, {Quantum Monte Carlo methods for nuclear
  physics}, Reviews of Modern Physics 87~(3) (2015) 1067--1118.
\newblock \href {http://arxiv.org/abs/1412.3081} {\path{arXiv:1412.3081}},
  \href {https://doi.org/10.1103/RevModPhys.87.1067}
  {\path{doi:10.1103/RevModPhys.87.1067}}.

\bibitem{2019PhRvC.100d5803L}
D.~{Logoteta}, {Consistent nuclear matter calculations with local three-nucleon
  interactions}, \prc 100~(4) (2019) 045803.
\newblock \href {https://doi.org/10.1103/PhysRevC.100.045803}
  {\path{doi:10.1103/PhysRevC.100.045803}}.

\bibitem{Fortin_PRC_2016}
M.~Fortin, C.~Provid{\^e}ncia, A.~R. Raduta, F.~Gulminelli, J.~L. Zdunik,
  P.~Haensel, M.~Bejger,
  \href{https://link.aps.org/doi/10.1103/PhysRevC.94.035804}{Neutron star radii
  and crusts: Uncertainties and unified equations of state}, Phys. Rev. C 94
  (2016) 035804.
\newblock \href {https://doi.org/10.1103/PhysRevC.94.035804}
  {\path{doi:10.1103/PhysRevC.94.035804}}.
\newline\urlprefix\url{https://link.aps.org/doi/10.1103/PhysRevC.94.035804}

\bibitem{Chen_PRC_2007}
Y.~Chen, H.~Guo, Y.~Liu,
  \href{https://link.aps.org/doi/10.1103/PhysRevC.75.035806}{Neutrino
  scattering rates in neutron star matter with \ensuremath{\Delta} isobars},
  Phys. Rev. C 75 (2007) 035806.
\newblock \href {https://doi.org/10.1103/PhysRevC.75.035806}
  {\path{doi:10.1103/PhysRevC.75.035806}}.
\newline\urlprefix\url{https://link.aps.org/doi/10.1103/PhysRevC.75.035806}

\bibitem{Zhu_PRC_2016}
Z.-Y. Zhu, A.~Li, J.-N. Hu, H.~Sagawa,
  \href{https://link.aps.org/doi/10.1103/PhysRevC.94.045803}{$\mathrm{\ensuremath{\Delta}}(1232)$
  effects in density-dependent relativistic hartree-fock theory and neutron
  stars}, Phys. Rev. C 94 (2016) 045803.
\newblock \href {https://doi.org/10.1103/PhysRevC.94.045803}
  {\path{doi:10.1103/PhysRevC.94.045803}}.
\newline\urlprefix\url{https://link.aps.org/doi/10.1103/PhysRevC.94.045803}

\bibitem{Sahoo_PRC_2018}
H.~S. Sahoo, G.~Mitra, R.~Mishra, P.~K. Panda, B.-A. Li,
  \href{https://link.aps.org/doi/10.1103/PhysRevC.98.045801}{Neutron star
  matter with $\mathrm{\ensuremath{\Delta}}$ isobars in a relativistic quark
  model}, Phys. Rev. C 98 (2018) 045801.
\newblock \href {https://doi.org/10.1103/PhysRevC.98.045801}
  {\path{doi:10.1103/PhysRevC.98.045801}}.
\newline\urlprefix\url{https://link.aps.org/doi/10.1103/PhysRevC.98.045801}

\bibitem{Kolomeitsev_NPA_2017}
E.~Kolomeitsev, K.~Maslov, D.~Voskresensky,
  \href{http://www.sciencedirect.com/science/article/pii/S0375947417300295}{Delta
  isobars in relativistic mean-field models with \ensuremath{\sigma}-scaled
  hadron masses and couplings}, Nuclear Physics A 961 (2017) 106 -- 141.
\newblock \href
  {https://doi.org/https://doi.org/10.1016/j.nuclphysa.2017.02.004}
  {\path{doi:https://doi.org/10.1016/j.nuclphysa.2017.02.004}}.
\newline\urlprefix\url{http://www.sciencedirect.com/science/article/pii/S0375947417300295}

\bibitem{Li_PLB_2018}
J.~J. Li, A.~Sedrakian, F.~Weber,
  \href{http://www.sciencedirect.com/science/article/pii/S0370269318305070}{Competition
  between delta isobars and hyperons and properties of compact stars}, Phys.
  Lett. B 783 (2018) 234 -- 240.
\newblock \href
  {https://doi.org/https://doi.org/10.1016/j.physletb.2018.06.051}
  {\path{doi:https://doi.org/10.1016/j.physletb.2018.06.051}}.
\newline\urlprefix\url{http://www.sciencedirect.com/science/article/pii/S0370269318305070}

\bibitem{Li2019ApJ}
J.~J. {Li}, A.~{Sedrakian}, {Implications from GW170817 for
  {\ensuremath{\Delta}}-isobar Admixed Hypernuclear Compact Stars}, \apjl
  874~(2) (2019) L22.
\newblock \href {http://arxiv.org/abs/1904.02006} {\path{arXiv:1904.02006}},
  \href {https://doi.org/10.3847/2041-8213/ab1090}
  {\path{doi:10.3847/2041-8213/ab1090}}.

\bibitem{Ribes_2019}
P.~Ribes, A.~Ramos, L.~Tolos, C.~Gonzalez-Boquera, M.~Centelles, {Interplay
  between $\Delta$ Particles and Hyperons in Neutron Stars}, ApJ 883 (2019)
  168.
\newblock \href {http://arxiv.org/abs/1907.08583} {\path{arXiv:1907.08583}},
  \href {https://doi.org/10.3847/1538-4357/ab3a93}
  {\path{doi:10.3847/1538-4357/ab3a93}}.

\bibitem{Li2020PhRvD}
J.~J. {Li}, A.~{Sedrakian}, M.~{Alford}, {Relativistic hybrid stars with
  sequential first-order phase transitions and heavy-baryon envelopes}, \prd
  101~(6) (2020) 063022.
\newblock \href {https://doi.org/10.1103/PhysRevD.101.063022}
  {\path{doi:10.1103/PhysRevD.101.063022}}.

\bibitem{1991PhRvL..67.2414G}
N.~K. {Glendenning}, S.~A. {Moszkowski}, {Reconciliation of neutron-star masses
  and binding of the Lambda in hypernuclei}, \prl 67 (1991) 2414--1417.
\newblock \href {https://doi.org/10.1103/PhysRevLett.67.2414}
  {\path{doi:10.1103/PhysRevLett.67.2414}}.

\bibitem{2005PhRvC..71f4301T}
S.~{Typel}, {Relativistic model for nuclear matter and atomic nuclei with
  momentum-dependent self-energies}, \prc 71~(6) (2005) 064301.
\newblock \href {http://arxiv.org/abs/nucl-th/0501056}
  {\path{arXiv:nucl-th/0501056}}, \href
  {https://doi.org/10.1103/PhysRevC.71.064301}
  {\path{doi:10.1103/PhysRevC.71.064301}}.

\bibitem{2010PhRvC..81a5803T}
S.~{Typel}, G.~{R{\"o}pke}, T.~{Kl{\"a}hn}, D.~{Blaschke}, H.~H. {Wolter},
  {Composition and thermodynamics of nuclear matter with light clusters}, \prc
  81~(1) (2010) 015803.
\newblock \href {http://arxiv.org/abs/0908.2344} {\path{arXiv:0908.2344}},
  \href {https://doi.org/10.1103/PhysRevC.81.015803}
  {\path{doi:10.1103/PhysRevC.81.015803}}.

\bibitem{PhysRevC.66.024306}
T.~Nik\ifmmode \check{s}\else \v{s}\fi{}i\ifmmode~\acute{c}\else \'{c}\fi{},
  D.~Vretenar, P.~Finelli, P.~Ring,
  \href{https://link.aps.org/doi/10.1103/PhysRevC.66.024306}{Relativistic
  hartree-bogoliubov model with density-dependent meson-nucleon couplings},
  Phys. Rev. C 66 (2002) 024306.
\newblock \href {https://doi.org/10.1103/PhysRevC.66.024306}
  {\path{doi:10.1103/PhysRevC.66.024306}}.
\newline\urlprefix\url{https://link.aps.org/doi/10.1103/PhysRevC.66.024306}

\bibitem{2005PhRvC..71b4312L}
G.~A. {Lalazissis}, T.~{Nik{\v{s}}i{\'c}}, D.~{Vretenar}, P.~{Ring}, {New
  relativistic mean-field interaction with density-dependent meson-nucleon
  couplings}, \prc 71~(2) (2005) 024312.
\newblock \href {https://doi.org/10.1103/PhysRevC.71.024312}
  {\path{doi:10.1103/PhysRevC.71.024312}}.

\bibitem{TANINAH2020135065}
A.~Taninah, S.~Agbemava, A.~Afanasjev, P.~Ring,
  \href{https://www.sciencedirect.com/science/article/pii/S0370269319307877}{Parametric
  correlations in energy density functionals}, Physics Letters B 800 (2020)
  135065.
\newblock \href
  {https://doi.org/https://doi.org/10.1016/j.physletb.2019.135065}
  {\path{doi:https://doi.org/10.1016/j.physletb.2019.135065}}.
\newline\urlprefix\url{https://www.sciencedirect.com/science/article/pii/S0370269319307877}

\bibitem{1997PhRvC..55..540L}
G.~A. {Lalazissis}, J.~{K{\"o}nig}, P.~{Ring}, {New parametrization for the
  Lagrangian density of relativistic mean field theory}, \prc 55~(1) (1997)
  540--543.
\newblock \href {http://arxiv.org/abs/nucl-th/9607039}
  {\path{arXiv:nucl-th/9607039}}, \href
  {https://doi.org/10.1103/PhysRevC.55.540}
  {\path{doi:10.1103/PhysRevC.55.540}}.

\bibitem{MATSUI1981365}
T.~Matsui,
  \href{https://www.sciencedirect.com/science/article/pii/0375947481901032}{Fermi-liquid
  properties of nuclear matter in a relativistic mean-field theory}, Nuclear
  Physics A 370~(3) (1981) 365--388.
\newblock \href {https://doi.org/https://doi.org/10.1016/0375-9474(81)90103-2}
  {\path{doi:https://doi.org/10.1016/0375-9474(81)90103-2}}.
\newline\urlprefix\url{https://www.sciencedirect.com/science/article/pii/0375947481901032}

\bibitem{PhysRevC.90.044305}
W.-C. Chen, J.~Piekarewicz,
  \href{https://link.aps.org/doi/10.1103/PhysRevC.90.044305}{Building
  relativistic mean field models for finite nuclei and neutron stars}, Phys.
  Rev. C 90 (2014) 044305.
\newblock \href {https://doi.org/10.1103/PhysRevC.90.044305}
  {\path{doi:10.1103/PhysRevC.90.044305}}.
\newline\urlprefix\url{https://link.aps.org/doi/10.1103/PhysRevC.90.044305}

\bibitem{2009PhRvL.102l2701T}
M.~B. {Tsang}, Y.~{Zhang}, P.~{Danielewicz}, M.~{Famiano}, Z.~{Li}, W.~G.
  {Lynch}, A.~W. {Steiner}, {Constraints on the Density Dependence of the
  Symmetry Energy}, \prl 102~(12) (2009) 122701.
\newblock \href {http://arxiv.org/abs/0811.3107} {\path{arXiv:0811.3107}},
  \href {https://doi.org/10.1103/PhysRevLett.102.122701}
  {\path{doi:10.1103/PhysRevLett.102.122701}}.

\bibitem{TSANG2011400}
M.~Tsang, Z.~Chajecki, D.~Coupland, P.~Danielewicz, F.~Famiano, R.~Hodges,
  M.~Kilburn, F.~Lu, W.~Lynch, J.~Winkelbauer, M.~Youngs, Y.~Zhang,
  \href{https://www.sciencedirect.com/science/article/pii/S0146641011000421}{Constraints
  on the density dependence of the symmetry energy from heavy-ion collisions},
  Progress in Particle and Nuclear Physics 66~(2) (2011) 400--404, particle and
  Nuclear Astrophysics.
\newblock \href {https://doi.org/https://doi.org/10.1016/j.ppnp.2011.01.041}
  {\path{doi:https://doi.org/10.1016/j.ppnp.2011.01.041}}.
\newline\urlprefix\url{https://www.sciencedirect.com/science/article/pii/S0146641011000421}

\bibitem{2020ApJ...896L..44A}
R.~{Abbott}, T.~D. {Abbott}, S.~{Abraham}, F.~{Acernese}, K.~{Ackley},
  C.~{Adams}, R.~X. {Adhikari}, V.~B. {Adya}, C.~{Affeldt}, M.~{Agathos},
  et~al., {GW190814: Gravitational Waves from the Coalescence of a 23 Solar
  Mass Black Hole with a 2.6 Solar Mass Compact Object}, \apjl 896~(2) (2020)
  L44.
\newblock \href {http://arxiv.org/abs/2006.12611} {\path{arXiv:2006.12611}},
  \href {https://doi.org/10.3847/2041-8213/ab960f}
  {\path{doi:10.3847/2041-8213/ab960f}}.

\bibitem{2021arXiv210506980R}
T.~E. {Riley}, A.~L. {Watts}, P.~S. {Ray}, S.~{Bogdanov}, S.~{Guillot}, S.~M.
  {Morsink}, et~al., {A NICER View of the Massive Pulsar PSR J0740+6620
  Informed by Radio Timing and XMM-Newton Spectroscopy}, arXiv e-prints (2021)
  arXiv:2105.06980\href {http://arxiv.org/abs/2105.06980}
  {\path{arXiv:2105.06980}}.

\bibitem{2021arXiv210506979M}
M.~C. {Miller}, F.~K. {Lamb}, A.~J. {Dittmann}, S.~{Bogdanov},
  Z.~{Arzoumanian}, K.~C. {Gendreau}, S.~{Guillot}, W.~C.~G. {Ho}, J.~M.
  {Lattimer}, M.~{Loewenstein}, S.~M. {Morsink}, P.~S. {Ray}, M.~T. {Wolff},
  C.~L. {Baker}, T.~{Cazeau}, S.~{Manthripragada}, C.~B. {Markwardt},
  T.~{Okajima}, S.~{Pollard}, I.~{Cognard}, H.~T. {Cromartie}, E.~{Fonseca},
  L.~{Guillemot}, M.~{Kerr}, A.~{Parthasarathy}, T.~T. {Pennucci}, S.~{Ransom},
  I.~{Stairs}, {The Radius of PSR J0740+6620 from NICER and XMM-Newton Data},
  arXiv e-prints (2021) arXiv:2105.06979\href {http://arxiv.org/abs/2105.06979}
  {\path{arXiv:2105.06979}}.

\bibitem{2021ApJ...908L..46R}
R.~W. {Romani}, D.~{Kandel}, A.~V. {Filippenko}, T.~G. {Brink}, W.~{Zheng},
  {PSR J1810+1744: Companion Darkening and a Precise High Neutron Star Mass},
  \apjl 908~(2) (2021) L46.
\newblock \href {http://arxiv.org/abs/2101.09822} {\path{arXiv:2101.09822}},
  \href {https://doi.org/10.3847/2041-8213/abe2b4}
  {\path{doi:10.3847/2041-8213/abe2b4}}.

\bibitem{2013Sci...340..448A}
J.~{Antoniadis}, P.~C.~C. {Freire}, N.~{Wex}, T.~M. {Tauris}, R.~S. {Lynch},
  M.~H. {van Kerkwijk}, et~al., {A Massive Pulsar in a Compact Relativistic
  Binary}, Science 340~(6131) (2013) 448.
\newblock \href {http://arxiv.org/abs/1304.6875} {\path{arXiv:1304.6875}},
  \href {https://doi.org/10.1126/science.1233232}
  {\path{doi:10.1126/science.1233232}}.

\bibitem{2019ApJ...887L..24M}
M.~C. {Miller}, F.~K. {Lamb}, A.~J. {Dittmann}, S.~{Bogdanov},
  Z.~{Arzoumanian}, K.~C. {Gendreau}, et~al., {PSR J0030+0451 Mass and Radius
  from NICER Data and Implications for the Properties of Neutron Star Matter},
  \apjl 887~(1) (2019) L24.
\newblock \href {http://arxiv.org/abs/1912.05705} {\path{arXiv:1912.05705}},
  \href {https://doi.org/10.3847/2041-8213/ab50c5}
  {\path{doi:10.3847/2041-8213/ab50c5}}.

\bibitem{2019ApJ...887L..21R}
T.~E. {Riley}, A.~L. {Watts}, S.~{Bogdanov}, P.~S. {Ray}, R.~M. {Ludlam},
  S.~{Guillot}, Z.~{Arzoumanian}, C.~L. {Baker}, A.~V. {Bilous},
  D.~{Chakrabarty}, K.~C. {Gendreau}, A.~K. {Harding}, W.~C.~G. {Ho}, J.~M.
  {Lattimer}, S.~M. {Morsink}, T.~E. {Strohmayer}, {A NICER View of PSR
  J0030+0451: Millisecond Pulsar Parameter Estimation}, \apjl 887~(1) (2019)
  L21.
\newblock \href {http://arxiv.org/abs/1912.05702} {\path{arXiv:1912.05702}},
  \href {https://doi.org/10.3847/2041-8213/ab481c}
  {\path{doi:10.3847/2041-8213/ab481c}}.

\bibitem{1971ApJ...170..299B}
G.~{Baym}, C.~{Pethick}, P.~{Sutherland}, {The Ground State of Matter at High
  Densities: Equation of State and Stellar Models}, \apj 170 (1971) 299.
\newblock \href {https://doi.org/10.1086/151216} {\path{doi:10.1086/151216}}.

\bibitem{1971NuPhA.175..225B}
G.~{Baym}, H.~A. {Bethe}, C.~J. {Pethick}, {Neutron star matter}, \nphysa
  175~(2) (1971) 225--271.
\newblock \href {https://doi.org/10.1016/0375-9474(71)90281-8}
  {\path{doi:10.1016/0375-9474(71)90281-8}}.

\bibitem{1996NuPhA.606..508M}
H.~{M{\"u}ller}, B.~D. {Serot}, {Relativistic mean-field theory and the
  high-density nuclear equation of state}, \nphysa 606 (1996) 508--537.
\newblock \href {http://arxiv.org/abs/nucl-th/9603037}
  {\path{arXiv:nucl-th/9603037}}, \href
  {https://doi.org/10.1016/0375-9474(96)00187-X}
  {\path{doi:10.1016/0375-9474(96)00187-X}}.

\bibitem{2001PhRvL..86.5647H}
C.~J. {Horowitz}, J.~{Piekarewicz}, {Neutron Star Structure and the Neutron
  Radius of $^{208}$Pb}, \prl 86~(25) (2001) 5647--5650.
\newblock \href {http://arxiv.org/abs/astro-ph/0010227}
  {\path{arXiv:astro-ph/0010227}}, \href
  {https://doi.org/10.1103/PhysRevLett.86.5647}
  {\path{doi:10.1103/PhysRevLett.86.5647}}.

\bibitem{PhysRevC.74.035802}
T.~Kl\"ahn, D.~Blaschke, S.~Typel, E.~N.~E. van Dalen, A.~Faessler, C.~Fuchs,
  T.~Gaitanos, H.~Grigorian, A.~Ho, E.~E. Kolomeitsev, M.~C. Miller,
  G.~R\"opke, J.~Tr\"umper, D.~N. Voskresensky, F.~Weber, H.~H. Wolter,
  \href{https://link.aps.org/doi/10.1103/PhysRevC.74.035802}{Constraints on the
  high-density nuclear equation of state from the phenomenology of compact
  stars and heavy-ion collisions}, Phys. Rev. C 74 (2006) 035802.
\newblock \href {https://doi.org/10.1103/PhysRevC.74.035802}
  {\path{doi:10.1103/PhysRevC.74.035802}}.
\newline\urlprefix\url{https://link.aps.org/doi/10.1103/PhysRevC.74.035802}

\bibitem{PhysRevC.84.065810}
R.~Cavagnoli, D.~P. Menezes, C.~m.~c. Provid\^encia,
  \href{https://link.aps.org/doi/10.1103/PhysRevC.84.065810}{Neutron star
  properties and the symmetry energy}, Phys. Rev. C 84 (2011) 065810.
\newblock \href {https://doi.org/10.1103/PhysRevC.84.065810}
  {\path{doi:10.1103/PhysRevC.84.065810}}.
\newline\urlprefix\url{https://link.aps.org/doi/10.1103/PhysRevC.84.065810}

\bibitem{2014EPJA...50...44P}
C.~{Provid{\^e}ncia}, S.~S. {Avancini}, R.~{Cavagnoli}, S.~{Chiacchiera},
  C.~{Ducoin}, F.~{Grill}, J.~{Margueron}, D.~P. {Menezes}, A.~{Rabhi},
  I.~{Vida{\~n}a}, {Imprint of the symmetry energy on the inner crust and
  strangeness content of neutron stars}, European Physical Journal A 50 (2014)
  44.
\newblock \href {http://arxiv.org/abs/1307.1436} {\path{arXiv:1307.1436}},
  \href {https://doi.org/10.1140/epja/i2014-14044-7}
  {\path{doi:10.1140/epja/i2014-14044-7}}.

\bibitem{CREX2021a}
{CREX Collaboration},
  \href{https://absimage.aps.org/image/APR21/MWS_APR21-2021-001167.pdf}{{}},
  APR21 Meeting (APS) (Apr. 2021).
\newline\urlprefix\url{https://absimage.aps.org/image/APR21/MWS_APR21-2021-001167.pdf}

\bibitem{CREX2021b}
\href{http://www2.latech.edu/~rakithab/project/rex/}{{}},
  http://www2.latech.edu/~rakithab/project/rex (Apr. 2021).
\newline\urlprefix\url{http://www2.latech.edu/~rakithab/project/rex/}

\end{thebibliography}





\end{document}